\documentstyle[twocolumn,epsf,prl,aps]{revtex}
\begin{document}
\draft

\twocolumn[
\hsize\textwidth\columnwidth\hsize\csname @twocolumnfalse\endcsname

\title{Integer filling metal insulator transitions in the
degenerate Hubbard model
}

\author{Marcelo J. Rozenberg}
\address{Laboratoire de Physique Th\'eorique de l'Ecole Normale 
Sup\'erieure,\\
24, rue Lhomond, 75231 Paris Cedex 05, France.\\
e-mail: marcelo@physique.ens.fr
}
\date{\today}

\maketitle
\widetext
\advance\leftskip by 57pt
\advance\rightskip by 57pt

\begin{abstract}
We obtain exact numerical solutions
of the degenerate Hubbard model in the limit of large
dimensions (or large lattice connectivity).
Successive Mott-Hubbard metal insulator transitions at
integer fillings  occur at intermediate values of the
interaction and low enough temperature in the paramagnetic phase.
The results are relevant for transition metal oxides 
with partially filled narrow degenerate bands.
\end{abstract}


]

\narrowtext

The understanding of strongly correlated electron systems is
one of the current challenges in condensed matter physics.
In recent years, this problem has received a great deal of
attention from theorists and experimentalist alike.
Yet, our knowledge of even some of the basic features of the proposed
model Hamiltonians remains to a large extent only partial,
except, perhaps, in the one-dimensional case.
In consequence, the interpretation of the experimental data
of strongly correlated electron systems has to remain only speculative
in most of the cases.
In regard of this situation, exact results on properties
of model Hamiltonians in well defined limits is very
desirable.
The relevant role played by the band degeneracy in models of
strongly correlated electrons has been long and 
largely recognized \cite{kanamori,nozieres,kugel,cyrot}.
However, the systematic treatment of such models poses in general
even greater technical difficulties than non-degenerate ones.
This particularly applies to numerical approaches which have to
deal with the exponential grow of the Hilbert space.

The goal of this paper 
is to demonstrate the existence of successive
metal insulator transitions (MIT) at the integer fillings $n=1, 2, 3$
in the two band degenerate
Hubbard model within the ``Local impurity self-consistent approximation''
(LISA) \cite{review} 
which is exact in the limit of large dimensions (or large
lattice connectivity) \cite{vollhardt}.
These MIT occur
within the paramagnetic phase for
intermediate values of the interaction.
We obtain exact numerical solutions of the 
model using an extension 
of the Hirsch and Fye quantum Monte Carlo (HFQMC) 
algorithm \cite{hirshfye,takegahara} 
for the
solution of the associated impurity problem within the LISA.
This impurity problem is a generalized single impurity Anderson
model where the impurity 
and conduction band operators carry an orbital index.

The two band degenerate Hubbard model reads,

\begin{eqnarray}
H  =  && \sum_{<ij>,a,b} t_{ij}^{ab} c_{ia\sigma}^\dagger c_{jb\sigma} +
  {(U+J) \over 2}  \sum_{i, a, \sigma } n_{ia \sigma} n_{ia -\sigma}
\nonumber\\
&& + {U \over 2} \sum_{i, a  \neq b, \sigma } n_{ia \sigma} n_{ib -\sigma}
 + {(U-J) \over 2}  \sum_{i, a  \neq b,  \sigma } n_{ia \sigma} n_{ib \sigma}
\nonumber\\
&& - {J \over 2} \sum_{i, a  \neq b, \sigma } c_{ia \sigma}^\dagger
c_{ia -\sigma} c_{ib -\sigma}^\dagger c_{ib \sigma}
\label{hamil}
\end{eqnarray}
$\langle ij \rangle$ labels nearest neighbor sites
and $a,b = 1,2$ is the orbital index.
This Hamiltonian is rotational invariant in spin and real space and the
usual approximation $U$ and $J$ independent of band indices
is assumed \cite{dworin}.
The parameter $U$ is due to on-site Coulomb repulsion and the exchange
parameter $J$ originates the Hund's coupling.
For simplicity, we shall further assume $t_{ij}^{ab}=-t\delta_{ab}$ and
neglect the last ``spin flip'' term in (\ref{hamil}).
The resulting model Hamiltonian
is relevant for for electronic systems with
partially filled narrow degenerate bands.
Examples of such systems are the $3d$ 
transition metal oxides $R_{1-x}A_xMO_3$
with three-dimensional perovskite-type structure, 
where $R=La,Y$, $A=Ca,Sr$ and
the transition metal $M=Ti,V,Cr$.
Fujimori has constructed a phase diagram by classifying these
compounds as correlated metals or Mott insulators
according to their electronic properties \cite{fujimori}.
Another and particularly notable example is the extensively 
investigated $V_2O_3$ compound which has a MIT as a function of temperature,
pressure and chemical substitution \cite{mcwhan}.
It is important to note that in bipartite lattices such as
the hypercubic or the Bethe lattice the groundstate of the model
has long range order \cite{kugel}.
In the single band case it is a spin antiferromagnetic (N\'eel)
state, while in the degenerate model
(with $J\approx 0$) one can
have either a spin ordered
and/or an orbital ordered state.
Other types of orderings are also a priori possible,
as for instance, ferromagnetic (for large $J$).
The stability of the different phases will depend
on the values of the various interaction parameters appearing in
the Hamiltonian (\ref{hamil}). Since the parameter
space is rather large, we defer the investigation
of the full phase diagram for a later publication.
Here we choose to concentrate on the interesting question
of the destruction of a metallic state by the sole effect of
electronic correlations (Mott transition). 
Therefore, we shall restrict ourselves
to the paramagnetic phase (in both spin and orbit indices).
It is important to note, however, that the paramagnetic solutions that
we obtain are
indeed the true groundstate in models
that include next nearest neighbor hopping (that lifts the rather
artificial nesting property) and also in models
with hopping disorder \cite{rkz,gekr}.

In this paper we shall consider the model on a Bethe lattice of
connectivity $z \rightarrow \infty$ which renders a semicircular
bare density of states 
$\rho^o (\epsilon) =
(2 / {\pi D}) \sqrt{1 - ( \epsilon /D)^2}$,
with $D=2t$.
Our choice is motivated from the fact that $\rho^o (\epsilon)$ shares
common features with a realistic three dimensional cubic tight binding model,
namely, it is bounded and has the same square root behavior at the edges.
We shall set the half-bandwidth $D=1$ and J=0 (the case
$J \ne 0$ will be considered elsewhere).

Within LISA,
the lattice model is exactly
mapped onto its associated impurity problem supplemented with a
self-consistency condition \cite{review}.
This approximation becomes exact in the limit of the dimensionality
$d \rightarrow \infty$
(or connectivity $z \rightarrow \infty$).
The resulting effective local action at a particular (any) lattice site
reads, \cite{gks,review}

\begin{eqnarray}
S_{loc}  =  && \int_0^\beta \int_0^\beta d\tau d\tau'
\sum_{<ij>,a,\sigma}
c_{a\sigma}^\dagger(\tau) {{\cal G}_{a\sigma}^0}^{-1}(\tau-\tau')
c_{a\sigma}(\tau') \nonumber \\
&& + {U \over 2} \int_0^\beta d\tau
\bigg( \sum_{i, a, \sigma } n_{ia \sigma} n_{ia -\sigma}
 + \sum_{i, a  \neq b, \sigma } n_{ia \sigma} n_{ib -\sigma}
\nonumber \\
&& + \sum_{i, a  \neq b, \sigma } n_{ia \sigma} n_{ib \sigma} \bigg)
\label{action}
\end{eqnarray}
and the Weiss field
${\cal G}_{a \sigma}^0$ is related to the local Green's function by the
self-consistency condition,

\begin{equation}
{{\cal G}_{a \sigma}^0}^{-1}(i\omega)
= i\omega + \mu - t^2  G_{a \sigma}(i\omega).
\label{consist}
\end{equation}

The LISA equations (\ref{action})-(\ref{consist}) have to be self-consistently
solved for.
In practice this is done by iteration. One begins with a guess 
for ${\cal G}_{a \sigma}^0$ which is used in (\ref{action}) to
solve the many-body problem and calculate $G_{a \sigma}$. This is
then used as input in (\ref{consist}) to get a new ${\cal G}_{a \sigma}^0$
and the process is iterated until self-consistency is attained.  
Eq.\ref{action} also defines an {\it impurity problem} where
the local site is coupled to an effective ``conduction band''
with hybridization function
$\Delta(i\omega) \equiv t^2  G_{a \sigma}(i\omega)$
similarly to the Anderson single impurity problem. This is the associated 
impurity problem within the LISA which has to be solved anew at
each step of the iteration.
To solve this {\it degenerate} impurity problem we use an extension of
the HFQMC algorithm. The
details for the implementation
can be found in Ref.\onlinecite{takegahara}.
One of its salient features is the absence of the negative sign problem
which allows the investigation of low temperatures ($\sim 10^{-2}$
of the bandwidth).  
Converged solutions are typically obtained after 8 iterations using
$L=64$ time slices in a few hours on a workstation.

\begin{figure}
\centerline{\epsfxsize=2.8truein
\epsffile{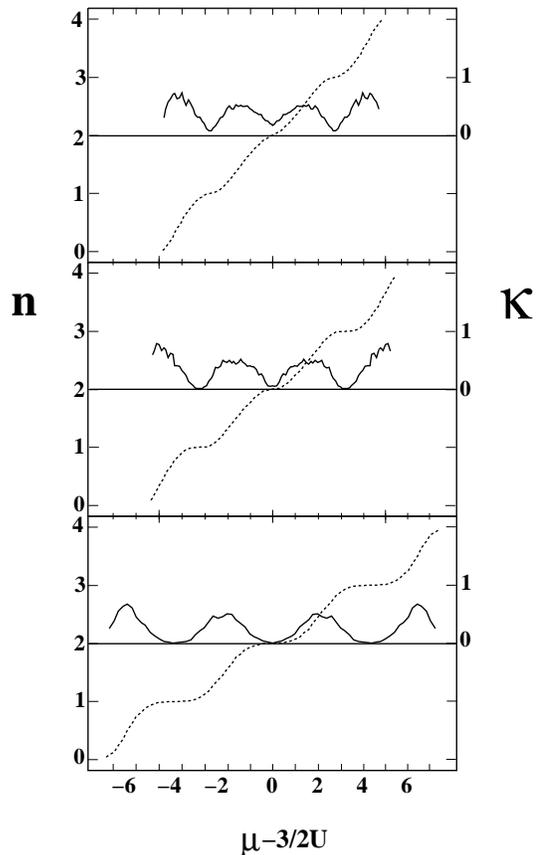}}
\vskip 0.1cm

\caption{
Occupation number $n$ and $\kappa = dn/d\mu$
(dotted and solid curves)
as a function of the chemical potential $\mu$. $\kappa$ is obtained
by numerical differentiation. The data are obtained at $U=2.5,3,4$ and
$T=1/8$ (top to bottom).
Note: for better comparison $\mu$ is
measured respect the particle-hole
symmetry point at $\frac{3}{2}U$.
}
\label{fig1}
\end{figure}

\begin{figure}
\centerline{\epsfxsize=2.8truein
\epsffile{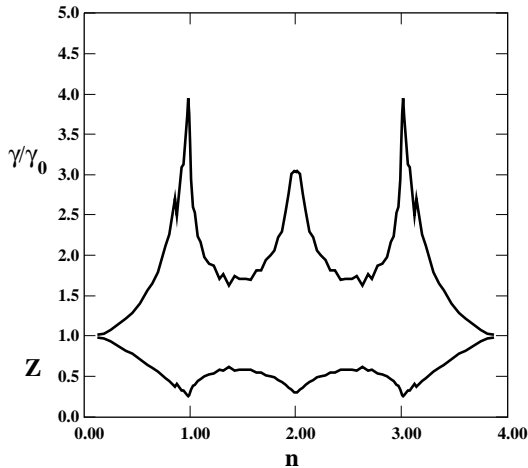}}
\vskip 0.1cm

\caption{
Quasiparticle residue $Z$ and 
specific heat $\gamma / \gamma_0$ (bottom and top)
as a function of the occupation number $n$ obtained
for $U=2.5$ and $T=1/8$. We estimate the
slope in the self-energy $\Sigma$ as the ratio of $\Sigma / \omega_n$
for the first Matsubara frequency. 
The wiggles are due to the QMC noise.
}
\label{fig2}
\end{figure}

We now turn to the  discussion of 
the results from the solution
of the LISA equations (\ref{action})-(\ref{consist}). In Fig.\ref{fig1}
we show the total occupation number $n=\langle
n_{1\uparrow}+n_{1\downarrow}+
n_{2\uparrow}+n_{2\downarrow}\rangle$
and and its first derivative $\kappa = dn/d\mu$ (which is 
proportional to the compressibility),
as a function of the chemical potential $\mu$.  
For $U=2.5$ we observe drops
in $\kappa$ at values of $\mu$ corresponding to
the integer fillings $n=1, 2, 3$ that indicates
the onset of a correlated metallic state. 
This is demonstrated in Fig.\ref{fig2} where we plot the quasiparticle
residue $Z = (1-\partial {\rm Im}\Sigma / \partial \omega_n)^{-1}$ 
and the specific 
heat $\gamma / \gamma_0$ ($= m^*/m_0$ in the $d= \infty$ limit). 
We observe that as the system approaches the integer fillings
the quasiparticles at the Fermi level become heavier as a consequence
of the proximity to the MIT. On the other hand, at $n=0$ and $4$ the
quasiparticle residue approaches unity as the correlations vanish.
It is worth noting the asymmetry in the behavior 
around $n=1$ and $3$. As one approaches $n=1$ from below the 
specific heat diverges as the inverse of the doping
$\delta^{-1}$ ($\delta=1-n$) \cite{kk}. 
However, as we approach $n=1$ from above 
the enhancement of $\gamma / \gamma_0$
seems to be faster.
The results for $0 \le n \le 1$ are similar 
as the obtained within a one band model and gives a posteriori justification
for its correct prediction of the
behavior of $\gamma / \gamma_0$ in 
$La_{1-x}Sr_xTiO_3$ \cite{rkz}.
It would be interesting to test the asymmetry prediction of our model
by approaching $n=1$ from above in oxigen deficient $LaTiO_{3-y}$.
Around $n=2$, which is the particle-hole symmetry point, 
the enhancement of $\gamma / \gamma_0$ is symmetric. 

Back to Fig.\ref{fig1}, we increase
the value of the interaction strength to $U=3$, and 
observe that the compressibility at $n=1$ and
$n=3$ vanishes as the system goes through a metal insulator 
transition,\cite{kk}
in fact, $U_c(n=1) = U_c(n=3)$ due to particle-hole
symmetry.
On the other hand, the compressibility at $n=2$ also decreases but
still remains finite, and the MIT
is reached upon further increasing the interaction.
Note that $\kappa$ also goes to zero at the endpoints 
$\mu-\frac{3}{2} \approx \pm 2U$, 
but these
are band insulating states which correspond to the 
completely empty $n=0$ and
completely full $n=4$ state.

\begin{figure}
\centerline{\epsfxsize=2.8truein
\epsffile{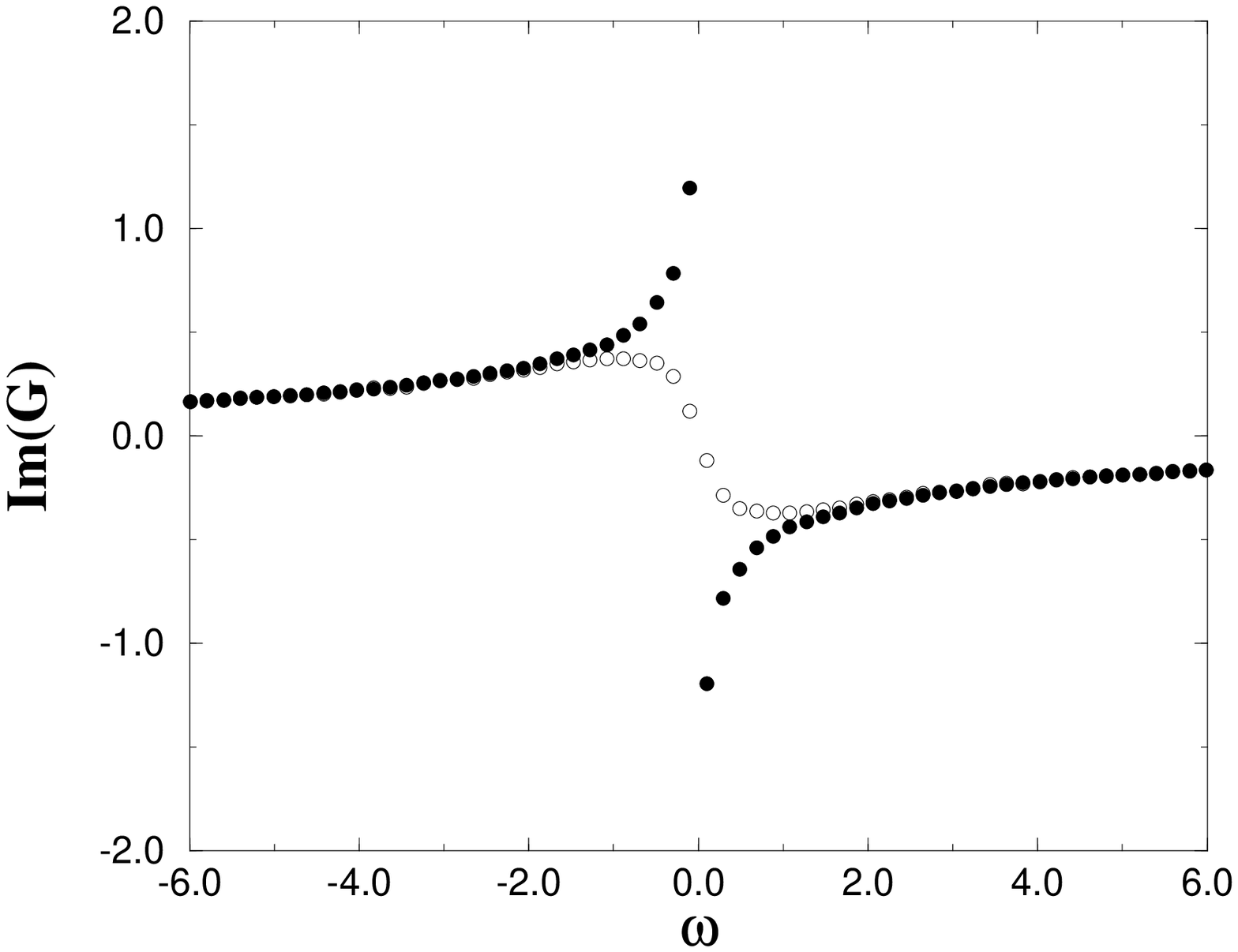}}
\centerline{\epsfxsize=2.8truein
\epsffile{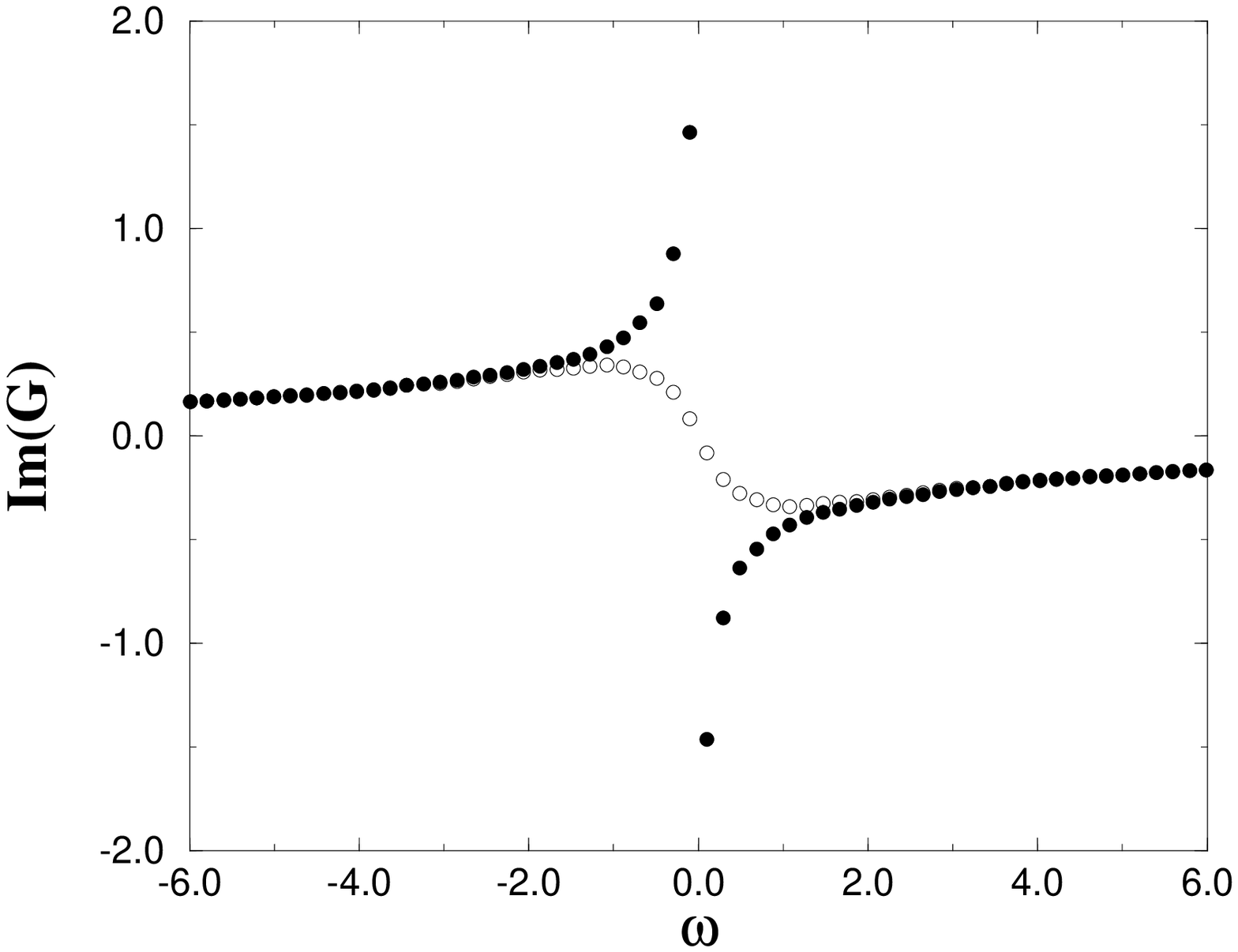}}

\caption{Top: Imaginary part of the Green's function as a function
of Matsubara frequency at $T=1/32$, $U=2.9$ and $n=1$. 
The solid dots correspond
to the metallic solution and the open dots to the insulating one.
Bottom: Idem for $T=1/32$, $U=3.4$ and $n=2$.}
\label{fig3}
\end{figure}

We find that the MIT of the degenerate model
at $n=1, 2, 3$ have
many features in common with the transition that occurs in the
single band case \cite{rkz}. In particular, at low
enough temperature 
$T \lesssim 0.04$ for $n=1,3$ and $T \lesssim 0.06$ for $n=2$, 
and within a finite interval of the interaction 
$U$ two
distinct solutions coexist: one has metallic character
while the other shows a gap at low frequencies.
In Fig.\ref{fig3} we show for the two fillings
$n=1$ and $2$
the coexistent Green's functions obtained
at $T=1/32$. 
To select either solution within the coexistent region, one has to
choose an appropriate ``seed'' at the beginning of the iteration.
These Green's functions are obtained as a function of Matsubara
frequencies, i.e., they live on the imaginary axis.
In order to obtain the density of states they have to be analytically
continued to the real axis.
Nevertheless, the metallic or insulating nature of the solution can
be unambiguously determined from their low frequency behavior.
In fact, the metallic Green's function behaves as ${\rm Im}G(\omega)
\ne 0$ for $\omega \rightarrow 0$ and can be continuously connected
to the non-interacting solution. On the other hand, the insulating
Green's function shows linear behavior at small frequencies
which is a signature of a gap opening in the density of states; thus, 
in this case,
${\rm Im}G(\omega) \rightarrow 0$ for $\omega \rightarrow 0$ and
can be continuously
connected to the atomic limit solution ($t \rightarrow 0$).

\begin{figure}
\centerline{\epsfxsize=2.8truein
\epsffile{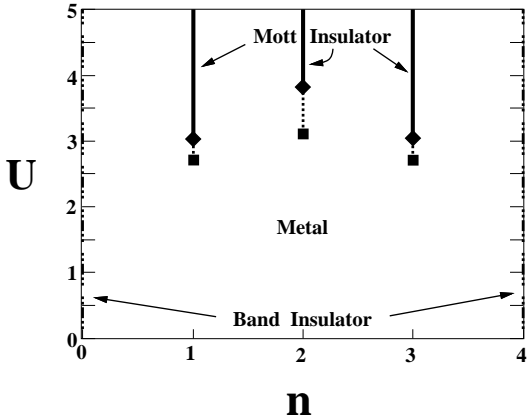}}

\vskip 0.1cm
\caption{Interaction $U$ vs. filling $n$
phase diagram for the paramagnetic solution.
The $1^{st}$ order transition lines at $n=1,2,3$ (dotted lines) 
end at $2^{nd}$ order
critical points. Diamonds indicate the approximate position of 
the $T=0$ critical points ($U_{c2}$), and
squares approximately indicate the 
finite $T$ critical point where the two solutions
merge ($T(n=1,3)\approx 0.04$, $T(n=2)\approx 0.06$). 
Solid lines at $n=1,2,3$ indicate the Mott insulator states, and
dashed-dotted lines at $n=0$ and $4$ 
indicate the empty and full band insulator states.
}
\label{fig4}
\end{figure}

Calling $U_{c1}(n)$ the minimum value of the
interaction for which an insulating
solution is allowed at given $n$, and $U_{c2}(n)$ the maximum value for
which a metallic solution is allowed;\cite{rkz}
we have established that at the lowest temperature investigated $T=1/32$,
$U_{c1}(1)=U_{c1}(3)\approx 2.8$, $U_{c2}(1)=U_{c2}(3)\approx 3$ and
$U_{c1}(2)\approx 3.1$, $U_{c2}(2)\approx 3.7$
Using similar arguments as for the single band model,\cite{moeller} 
one can demonstrate
that at $T=0$ the metallic solution is lower in energy, thus, the
$T=0$ MIT occurs at $U_{c2}$ and is of second order.
A direct consequence of two solutions being allowed in
regions of the ($U,T$) plane at given integer fillings, is the
existence of a first order transition line defined
by the crossing of their free
energies. This line starts at $U_{c2}$ at $T=0$ and 
has a negative slope due to the large entropy of the insulating state, 
while at finite $T$ it ends 
at another second order point where the two solutions
eventually merge. These results are condensed into the
phase diagram of Fig.\ref{fig4} which shows qualitative agreement with
that of Ref.\onlinecite{fujimori}.

{}From the experimental standpoint, it has been argued by 
Fujimori {\it et al.} \cite{fujimorietal} that the systematic
analysis of the electronic properties of various transition metal oxide
compounds with nominally {\it one electron} in a degenerate 
$d-$band shows
evidence for a MIT due to electron correlations.
Moreover, the extensively investigated $V_2O_3$ compound
contains nominally {\it two electrons} in a doubly degenerate $e_g$ band
and its phase diagram displays a paramagnetic MIT line which is {\it first 
order} \cite{mcwhan}.
Nevertheless, Castellani {\it et al.} \cite{castellani} have argued
that a relevant model for this system should contain just one electron
in a doubly degenerate $e_g$ band.
In either case 
the qualitative physics is being 
correctly predicted by our results.

In conclusion, we obtain
exact numerical solutions of the LISA mean field equations
for the degenerate Hubbard model. These are exact
solutions of the model in the limit of $d=\infty$.
We find that within the paramagnetic phase the
model has metal insulator transitions
at the integer fillings $n=1, 2, 3$ which are similar in character as the
single band case. In particular we find that
$U_c(n=2) > U_c(n=1) = U_c(n=3)$,
and that there are regions in the ($U,T$) plane
where two different solutions are allowed, 
one metallic and one insulating,
which leads to a first order MIT line at finite temperatures.
An important remark is that the observation of a similar scenario 
for the integer filling MIT in both, the single and degenerate band
model, validates the often made assumption about the relevance of
a single band Hubbard model for the qualitative investigation
of compounds with narrow degenerate bands. This remark can be considered
an a posteriori justification for the recent success in the interpretation
of low frequency spectroscopies of correlated systems with band
degeneracy using a single band model \cite{v2o3,inoue}. 
However, the previous statement has to be taken with care as we have seen
that some quantities, as for instance the specific heat, 
may show a different behavior depending how 
the MIT is approached. 

Many interesting questions are open ahead, for instance, 
whether the present scenario is modified by the introduction of a finite $J$
and the study of phases with long range order 
as function of $U,\ J,\ T$ and $n$. 
These are certainly relevant questions for
the understanding of the physics of
transition metal oxides with partially filled $d-$bands.

Valuable conversations with A. Georges, 
I. H. Inoue, H. Kajueter and G. Kotliar
are gratefully acknowledged.

\end{document}